Theoretical investigation on potential of zero free charge of (111) and (100) surfaces of Group 10 and 11 metals

Jack Jon Hinsch[1], Jessica Jein White, Yun Wang[1*]

1: Centre for Catalysis and Clean Energy, School of Environment and Science, Griffith University, Gold Coast, QLD 4222, Australia

* Corresponding Author's E-mail: yun.wang@griffith.edu.au

**Abstract**

The potential of zero free charge (PZFC) value is a crucial parameter in electrochemistry. However, the evaluations of PZFC have traditionally been difficult. To overcome this challenge, we applied a hybrid solvation method that incorporates, both an explicit water layer next to the metal surface and an implicit water layer, combined with density functional theory (DFT) to simplify the PZFC evaluation. Using the (111) and (100) surfaces of Group 10 and 11 metals as model systems, we calculated their PZFC values, which showed excellent agreement with the reported data. This great match validates the accuracy and reliability of our theoretical approach. Notably, we observed that the surface structure and the orientation of water molecules have a significant influence on the PZFC values of the metals. Our study, therefore, paves the way for efficiently and accurately calculating the PZFC values of materials, which can greatly benefit their practical applications.

**Highlights:**

- Implicit solvation insufficient to accurately predict the solid-liquid interface.
- Hybrid solvation techniques accurately predict potential of zero free charge.
- Morphologies of metal surfaces and water orientation greatly alter the potential of zero free change of Group 10 and 11 metals.

# 1. Introduction

Electrochemical redox reactions are of great interest in several prominent technologies, including electrocatalysis, batteries, supercapacitors, solar cells, and fuel cells [1-3]. The charge transfer across the electrified metal/water interface is one of the most important steps in these redox reactions, which is governed by applied electrode potential [4]. Given the difficulty that arises from the operando characterizations of the solid-liquid interfaces [5], it becomes paramount that a theoretical understanding of the applied electrode potentials is well defined [6, 7]. However, the electrode potential is difficult to theoretically control as finding an alignment scheme to compare well-defined electrode potentials proves challenging [4].

Several *ab-initio* simulation strategies have been developed thus far to characterize the electrified solid-liquid interface [8]. The computational hydrogen electrode (CHE) is widely used for studying the electrified metal/water interface [9, 10]. In this method, the standard hydrogen electrode (SHE) serves as the reference potential by introducing a pair of protons ($H^+$) and electrons ($e^-$) in each elementary step [11-13]. However, this method dismisses the influence of the applied bias potential on the properties of the solid-liquid interface [14]. To fill this gap, the potential of zero charge (PZC) has been proposed to act as the reference potential for theoretically manipulating the electrode potentials [8]. The PZC is an important parameter in electrochemistry that refers to the electric potential at which a solid surface becomes electrically neutral when immersed in an electrolyte solution [15, 16]. At this point, the electrode surface does not carry a net electrical charge, and the chemical potential of electrons on the surface is equal to that in the bulk of the solution. It plays a crucial role in understanding the surface properties of materials, including adsorption, corrosion, and catalysis [1]. Consequently, knowledge of the PZC is essential for the design and optimization of various electrochemical processes and devices [1-3].

Frumkin and Petri proposed two different PZCs to identify the fundamental magnitudes of the electrode-electrolyte interface [17]. The PZTC involves transferring the total charge, which includes the charge transferred during the chemisorption of species (e.g., hydrogen atoms). The PZFC, on the other hand, is observed in the absence of adsorption of species and is equivalent to the PZC for simplicity. The PZFC can be experimentally determined by measuring the differential capacitance of the electrode-electrolyte interface as a function of the applied potential [18]. It is important to note that experimental determination of the PZFC can be challenging due to factors such as impurities on the surface, surface roughness, and the presence of adsorbed species. Due to the structure of the Helmholtz planes, a strong electric field is produced, which causes further challenges when studying these interactions. As a powerful alternative, theoretical methods, such as density functional theory (DFT) calculations, have been used to predict and calculate the PZFC [2, 16]. This predictive capability is particularly useful when dealing with novel or complex materials where experimental measurements may be challenging or time-consuming. Moreover, theoretical calculations offer a deeper understanding of the electronic structure and charge distribution at the solid-liquid interface [19]. This information can shed light on surface properties, which are crucial for determining PZFC values.

Previously, the *ab-initio* molecular dynamics (AIMD) method is often employed to account for dynamic water structures [20]. Li et. al. investigated metal-water interactions aiming to understand the mechanisms of the interface water molecules on the PZC [15]. They found that the water molecules' orientation and proximity to the surface pushed the spilling electrons from the (111) surface back into the metal skeleton. The outer Helmholtz plane supported this transfer by holding the first layer in position. Bouzid et al. used the AIMD-based constant Fermi method to theoretically calculate the PZFC of electrified Pt(111)/water interface is 0.22 vs SHE [4]. Groß et al. conducted the AIMD to theoretically estimate the PZFC value of Pt(111) is 0.52 vs SHE [20]. While AIMD studies have proven effective in reproducing the experimental electrochemical properties of metallic interfaces, their computational cost poses significant limitations, particularly when applied to techniques like microkinetic modelling, where numerous elementary steps are involved in calculating complex reaction mechanisms [21]. A recent study conducted by Mathew et al. introduced an implicit solvation method for determining the electrode potential in a system consisting of an electrode and an electrolyte solution [22, 23]. They reported the PZC for the Pt(111) electrode to be 0.85 V. In their model, the solvent is treated as a continuous medium, considering its collective effects on the electrode-electrolyte interface [22, 23]. The solute is immersed in a solvent "bath," and the averaging over the solvent's degrees of freedom is incorporated into the simulation approach [24-27]. However, only using implicit solvation can prove misleading. It often underestimates electrochemical properties due to the absence of an atomistic description of the hydrogen bond, which is directional and short-range. It has been shown that the direction of the hydrogen bond can influence overall electrochemical properties [6]. Water orientation and chemisorption on metal electrodes can induce electronic redistribution, which changes the PZC [28, 29]. Therefore, the simulation using an implicit solvation model alone may not be sufficient.

In this study, we combine both methods into a cost-effective approach based on DFT and a double reference method to investigate the electrified interface. This hybrid approach allowed us to capture the essential features of the electrified interface while maintaining computational efficiency. To test the limitations of this model, Group 10 and 11 metals were studied. These metals were chosen as they are often relatively chemically inert. As a result, the impact of adsorption on these metal surfaces on the PZFC values can be small. The two (111) and (100) surfaces were purposely chosen since they are widely studied metal surfaces with dense surface structure and lower chemical reactivity in comparison with the more open metal surface, e.g., the (110) surface [30].

## 2. Computational Details

All simulations were performed within density functional theory using the Vienna Ab initio Simulation Package (VASP) [31] accompanied by the projected augmented wave (PAW) method [32]. Therefore, the electron – interactions were described via the PAW potentials. The valance electron configurations of Ag, Au, Ni, Pt, Pd and Cu are $5s^1\ 4d^{10}$, $6s^1\ 5d^9$, $4s^1\ 3d^9$, $6s^1\ 5d^9$, $4s^1\ 4d^9$, $4p^1\ 3d^{10}$, respectively. The H and O atoms had valance electron configurations of $1s^1$, and $2s^2\ 2p^4$, respectively. The spin-polarization was only considered when studying the Ni interface since it is magnetic. The wave function was expanded with a kinetic energy cut-off value of 600 eV [23]. The gamma-centered k-point meshes utilized a reciprocal space resolution of $2\pi \times 0.04$ Å$^{-1}$. The convergence criterion for the electronic self-consistent loop was set to $1 \times 10^{-5}$ eV. All atoms were allowed to relax till the Hellmann-Feynman forces were smaller than 0.05 eV/ Å. The exchange-correlation functional optPBE was used to study the electrified metal interface. The functional had been used before in our previous work and has produced reasonable results with platinum and water properties [33, 34]. This is because nonlocal correlation energy was added to the XC energy by using the optPBE with the less spherical densities accounted for to better incorporate the weaker interactions [35].

The (111) and (100) surfaces of six different Groups 10 and 11 metals, including Ni, Cu, Pd, Ag, Pt and Au, were investigated here using four-layer slab models. The hybrid solvation model was used to calculate the PZFC values of these surfaces. To facilitate the convergence of the electrostatic potential of the water solution, one layer of explicit water molecules was added on both the top and bottom sides of the metal slabs to improve the symmetry of the models. Three initial water configurations with different orientations were used. In the UP configuration, one hydrogen atom in half of the water molecules pointed away from the surface. In the DOWN configuration, these hydrogen atoms pointed to the surface, and a MIX configuration used an even distribution of UP and DOWN configurations. An implicit electrostatic continuum was used to simulate the impact of the rest part of the water solution with a permittivity of 78.40 and a Debye length of 3.040 Å [22, 23].

During zero charge calculations, the water molecules were allowed to relax. The potential of zero free charge (PZFC) was calculated with respect to the standard hydrogen electrode (SHE) using a double-reference method [23],

$$PZFC(V)\ vs.SHE\ =\ (-E_{shift} - E_{fermi})/|e| - 4.6 \qquad (1)$$

where the potential energy between the electrostatic potential energy of the implicit bulk electrolyte ($E_{shift}$) and Fermi energy of the system ($E_{fermi}$) were first calculated. After that, the applied bias potential with respect to the SHE potential was calculated by subtracting a constant of 4.6 V, which was proposed and validated previously by other groups [23, 36].

The distance between the surface and water layer is important for understanding interfacial properties. The distance was calculated by averaging the heights of all oxygen atoms in water and then comparing that to their respective surface. Since the metal surface was sandwiched between two water layers, the oxygen atoms needed to be compared to the closest metal surface. Therefore, the average M – O distance was calculated as:

$$d_{M-O}\ =\ \frac{(d_{Mtop}-d_{Otop})+(d_{Obot}-d_{Mbot})}{2} \qquad (2)$$

Where, $d_{Mtop}$ and $d_{Mbot}$ are the average z-position of all metal atoms on the top and bottom of the slab, respectively. $d_{Obot}$ and $d_{Otop}$ are the average z-position of all oxygen atoms below and above the metal slab, respectively.

## 3. Results and Discussion

### 3.1 Implicit solvation method

The PZFC values of the metal surface have first been calculated using the implicit solvation-only model. The atomic configurations of the metal (111) and (100) surfaces were shown in **Figs. 1a** and **b**. Both surfaces are flat. In the (111) surface, the coordination number (CN) of the surface atoms are 9. As a comparison, the CN of the (100) surface atoms are 8. As a result, the (111) surface of Group 10 and 11 metals are denser and more stable than the (100) surface. The comparison between our calculated PZFC values and the experimental ones seen in **Fig. 1c**. It was noticed that the PZFC values of the surfaces of Group 11 metals, including Au, Ag and Cu, using the implicit solvation model are close to the experimental values. This matches the previous conclusion of Mathew et al. [23] using the same model. The great match may be ascribed to the weak interaction between water and these surfaces since they are hydrophobic [37].

However, the theoretical PZFC values of the Group 10 metal surfaces show a significant deviation from the experimental data. The PZC value calculated for Pt(111) using the explicit method was found to be 1.09 V vs SHE. Our calculated Pt(111) PZFC value closely aligns with a similar work by Mathew et al. using the same implicit model with a value of 0.85 V [23]. However, greatly higher than the experimentally reported value of around 0.4 V documented in various studies [38-42] (see **Table S1**). The biggest deviations between our results and experimental values were found to be the Pd(100) and Pt(100) surfaces, which theoretical PZFC are about 0.6 V larger than the experimental data. Overall, the mean absolute error (MAE) of our calculated PZFC is 0.55, and the root mean squared error (RMSE) is 0.68. The considerably high errors support the conclusion made by Mathew et al. that the implicit solvation model alone is insufficient to accurately reproduce experimental PZFC values [23]. This is mainly ascribed to the ignorance of the impact of the explicit waters next to the surface.

Several groups have reported that the first layer of the water next to the metal surface has many different properties in terms of the bulk solvent [19, 29]. Moreover, the explicit water molecules also subtly impact the properties of the metal surfaces. As a result, it becomes imperative to build a hybrid approach that includes an explicit water layer next to the metal surface and an implicit solvent bulk to address this accuracy issue without compromising the computational cost.

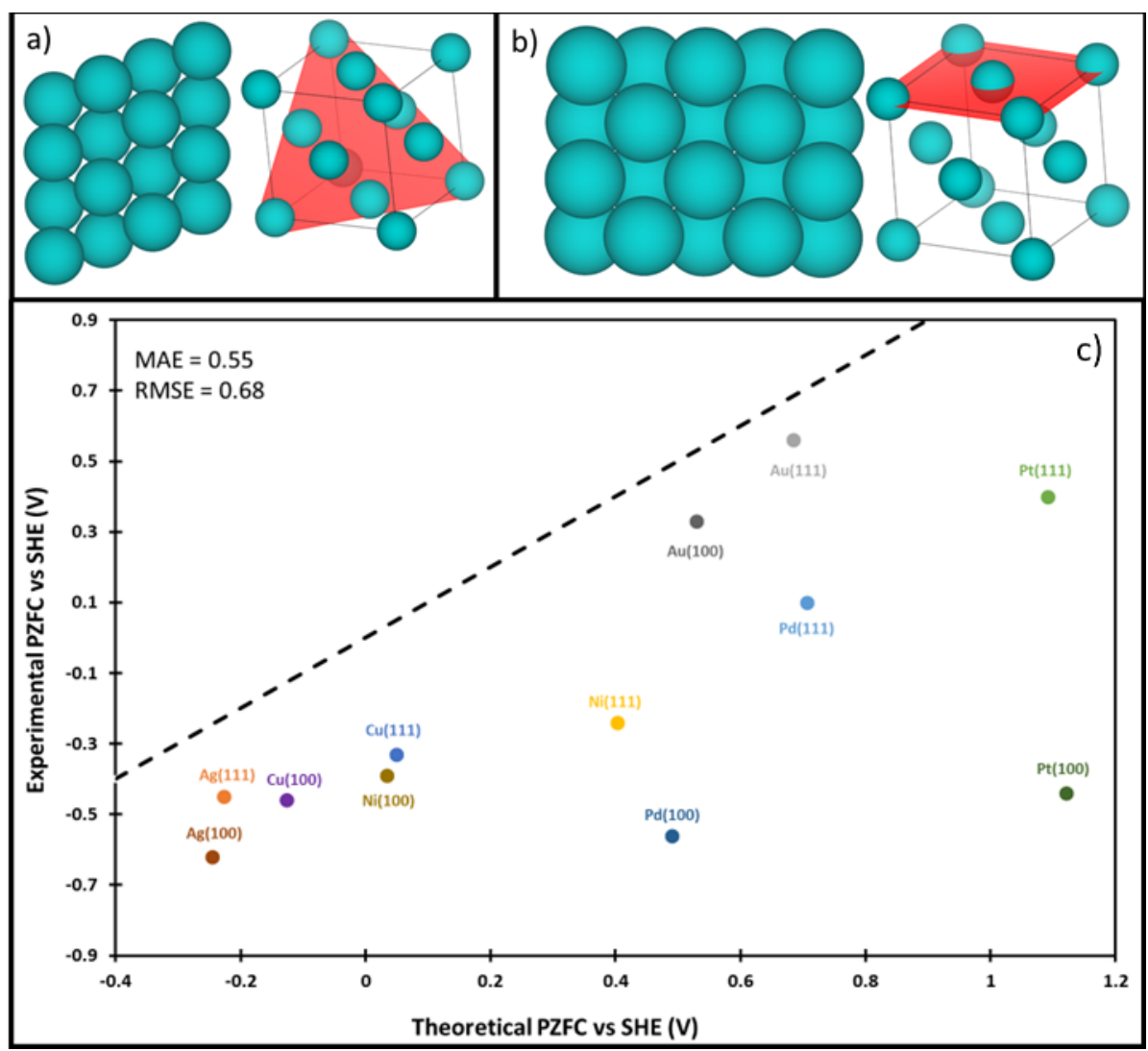

**Figure 1**: (a) The illustration of the metal (111) surface; (b) the illustration of the metal (100) surface; and (c) The comparison of our calculated PZFC values and the corresponding experimental data.

## 3.2 Hybrid solvation model

### 3.2.1 (111) Surface

The (111) surface is a fundamental flat surface and is often the least active of the Miller family for Group 10 and 11 *fcc* metals. The optimized atomic structure of the metal (111) surfaces with the hybrid solvation models are shown in **Fig. 2**. It can be found that the explicit water layers next to the (111) surface are still well uniform with distinct differences between UP, DOWN and MIX orientations. This phenomenon has been documented in previous works [4, 43]. The DOWN and MIX configuration often resulted in a smaller $d_{M-O}$ distance (see **Fig. 2**). An exception is the Cu(111) surface, which had a large $d_{Cu-O}$ distance when MIX orientation was used. Furthermore, Cu and Au had the largest interface distance regardless of orientation. Ni was noted to be visually disordered despite water orientation. Both Pd and Pt were similar, boasting slight deformations in the rigid ice layer while maintaining a short interfacial distance. The water layers on the Ag(111) and Au(111) surfaces had minimal deformation as both metals are hydrophobic and inert [37]. This also supports the larger distance seen on Group 11 elements. The formation of a stable water configuration is essential for creating the inner Helmholtz plane. Most metals formed an almost flat ice plane structure. Meanwhile, Ni showed a different construct. By varying the height of its water molecules, the shape of a slanted hexagon can fit in the restricting Ni lattice. The original hexagon seen on other metals was pinched at one corner, converting the shape into a chair-like layer reminiscent similar to that seen on graphene [44].

The orientation of the water layer played a strong role in how large the interface was. Generally, the UP configuration resulted in a larger $d_{M-O}$ distance than DOWN or MIX. Meanwhile, the DOWN configuration predominantly showed smaller $d_{M-O}$ distances. There were exceptions, like with Cu, which had a larger distance when the MIX orientation was used. However, this was presumed as this configuration was very disordered, which may lead to significant outliers shifting the average distance.

Our results demonstrate the importance of water orientations on the PZFC. For Group 11 elements, the MIX configuration was closest to experiments and the most stable starting configuration. Meanwhile, The impacts of Group 10 elements were split [45]. Pd preferred the MIX configuration, while Pt and Ni favored optimized into the DOWN orientation. Both DOWN and MIX configurations, in this case, were very similar, and it could be argued that they are the same structure. Additionally, across all metals, the $E_{total}$ difference between orientations was within only a few eV (see **Tables S2-3**). It suggests that there might be many local minimums during the interaction with distinct PZFC and interfacial properties, which matches the dynamic nature of the water liquid.

| | Cu EXP PZFC = -0.33 vs SHE (V) a) | Ag EXP PZFC = -0.45 vs SHE (V) a) | Au EXP PZFC = 0.56 vs SHE (V) a) |
|---|---|---|---|
| Down | 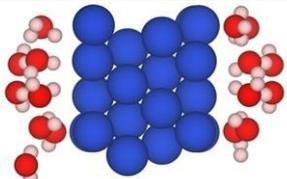<br>PZFC = 0.07 vs SHE (V) | 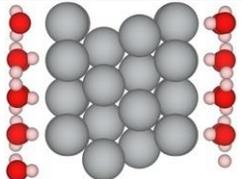<br>PZFC = 0.08 vs SHE (V) | 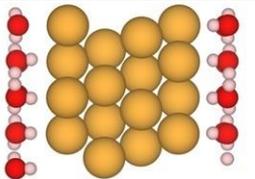<br>PZFC = 0.81 vs SHE (V) |
| Mix | 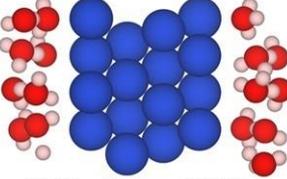<br>PZFC = -0.51 vs SHE (V) | 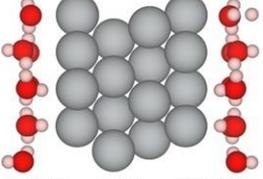<br>PZFC = -0.69 vs SHE (V) | 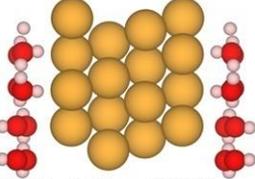<br>PZFC = 0.27 vs SHE (V) |
| Up | 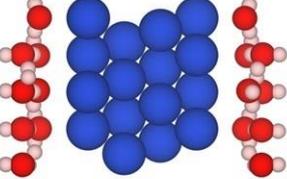<br>PZFC = -1.99 vs SHE (V) | 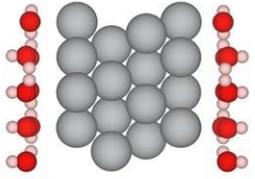<br>PZFC = -0.93 vs SHE (V) | 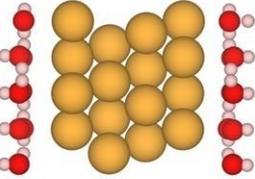<br>PZFC = -0.21 vs SHE (V) |
| | Ni EXP PZFC = -0.24 vs SHE (V) a) | Pd EXP PZFC = 0.1 vs SHE (V) b) | Pt EXP PZFC = 0.4 vs SHE (V) c) |
| Down | 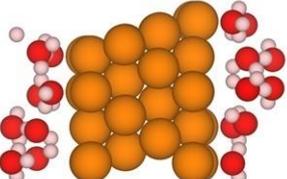<br>PZFC = -0.14 vs SHE (V) | 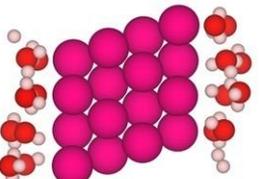<br>PZFC = 0.50 vs SHE (V) | 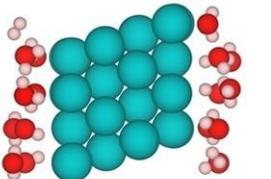<br>PZFC = 0.47 vs SHE (V) |
| Mix | 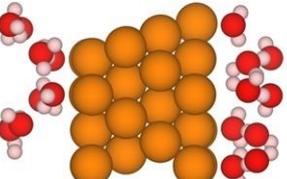<br>PZFC = -0.71 vs SHE (V) | 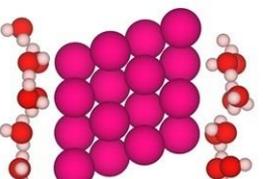<br>PZFC = -0.09 vs SHE (V) | 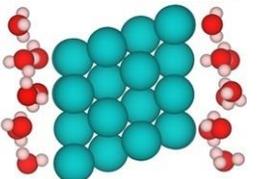<br>PZFC = 0.02 vs SHE (V) |
| Up | 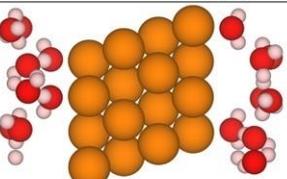<br>PZFC = -1.49 vs SHE (V) | 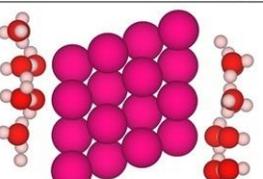<br>PZFC = -0.87 vs SHE (V) | 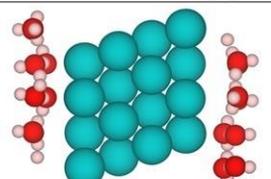<br>PZFC = -0.50 vs SHE (V) |

**Figure 2**. Metal water interface structures for (111) surfaces. Each interface has its PZFC below the image. On the top row of each element is the experimental approximation. Color legend: blue is Cu, silver is Ag, gold is Au, orange is Ni Pd is pink, Pt is green, red is O and pink is H.

a) from the reference [45]

b) from the reference [46]

c) from the references [38-42]

## 3.2.2 (100) Surface

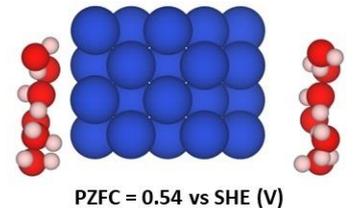

**Figure 3**. Metal water interface structures for (100) surfaces. Each interface has its PZFC below the image. On the top row of each element is the experimental approximation. Color legend: blue is Cu, silver is Ag, gold is Au, orange is Ni Pd is pink, Pt is green, red is O and pink is H.
a) from the reference [45].

The optimized metal (100)/water interfaces are shown in **Fig. 3**. In comparison with the configurations on the (111) surface, the (100) surface promoted the disordered water plane, which resulted in a large $d_{M-O}$ distance. Again, the largest M-O distance for all orientations was found on the Au(100) surface. However, the Au-O distance is exceptionally larger than that on the (111) surface. There is a 0.4 Å increase in their $d_{M-O}$ on the rest of the metal (100) surfaces. The orientation data also differed. On the (111) surface, the water orientations in DOWN and MIX have their own unique properties. On the (100) surface, both Cu and Ag MIX were converted into an exact configuration of the UP configuration after the structural optimizations. It indicates that the (100) surface has a higher impact on the water adsorption due to their high reactivities in terms to the (111) surface [30].

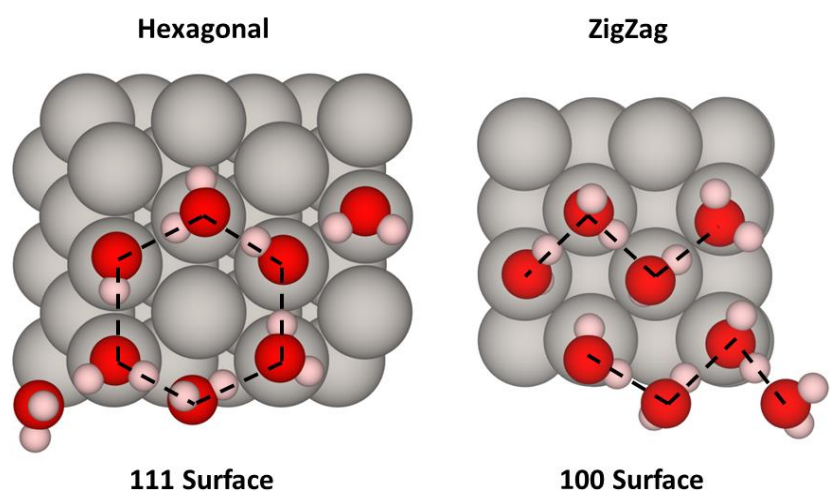

**Figure 4**. Varying water structures that form on the (111) and (100) surfaces. The structures above are seen on the (111) MIX and 100 DOWN structures respectively. Color legend: grey is metal Pt, red is O and pink is H.

The *fcc* (100) surface shows the orthogonal symmetry, which is different from the hexagonal symmetry of the ice crystal [20]. As a result, the water framework is distorted from the ice-like shape on the (100) surface. Previous study shows that water can form pentagons instead of hexagons on the (100) surface [47, 48]. In this work, a zigzag structure was found after the structural optimization on all (100) surfaces, which is close to the reported pentagon structures.

This deformation of water structures breaks some of the H-bonded networks, which swing away, forming the zigzag layer (see **Fig. 4**).

The impact of the water orientation on the largest $d_{M\text{-}O}$ at the water-metal (100) interfaces is not significant as that at the water-metal (111) interfaces. Ag, Au and Pt showed the UP orientation had the largest $d_{M-O}$, while Pd, Ni and Cu had a large UP orientation interfacial distance. The total energy gradual differences between all metals and orientations, with only at most 1 eV, except for Ni (see **Tables S4-5**). The MIX configuration on the Ni(100) surface had large total energy compared to the other configurations. This was believed to be because the MIX configuration had a clearly defined zigzag structure compared to other orientations. The preferred configuration varied based on the metal. Ag, Au, and Pt preferred the DOWN configuration. Meanwhile, Ni and Pd achieved their lowest total energy in the MIX configuration. Cu(100) was the only system that kept the UP configuration after the optimization.

### 3.3 Trends and Impacts

The PZFC is a critical parameter in interface studies, and its accurate determination is essential for understanding and predicting the behavior of materials in electrochemical processes. The results of the study demonstrate the reliability and accuracy of the hybrid solvation method for modeling the PZFC at solid-liquid interfaces. The method outperformed the implicit-only approach, with the theoretical PZFC values showing excellent agreement with established experimental data. The deviation between the calculated and experimental PZFC values using the hybrid approach was limited to within 0.3 eV, indicating the high accuracy of the method. Furthermore, the MAE (0.17) and RMSE (0.19) values were almost only 1/3 of those obtained with the implicit-only approach (0.55 and 0.68 for MAE and RMSE, respectively, see **Fig. 1c**), further confirming the superiority of the hybrid solvation method in predicting PZFC values.

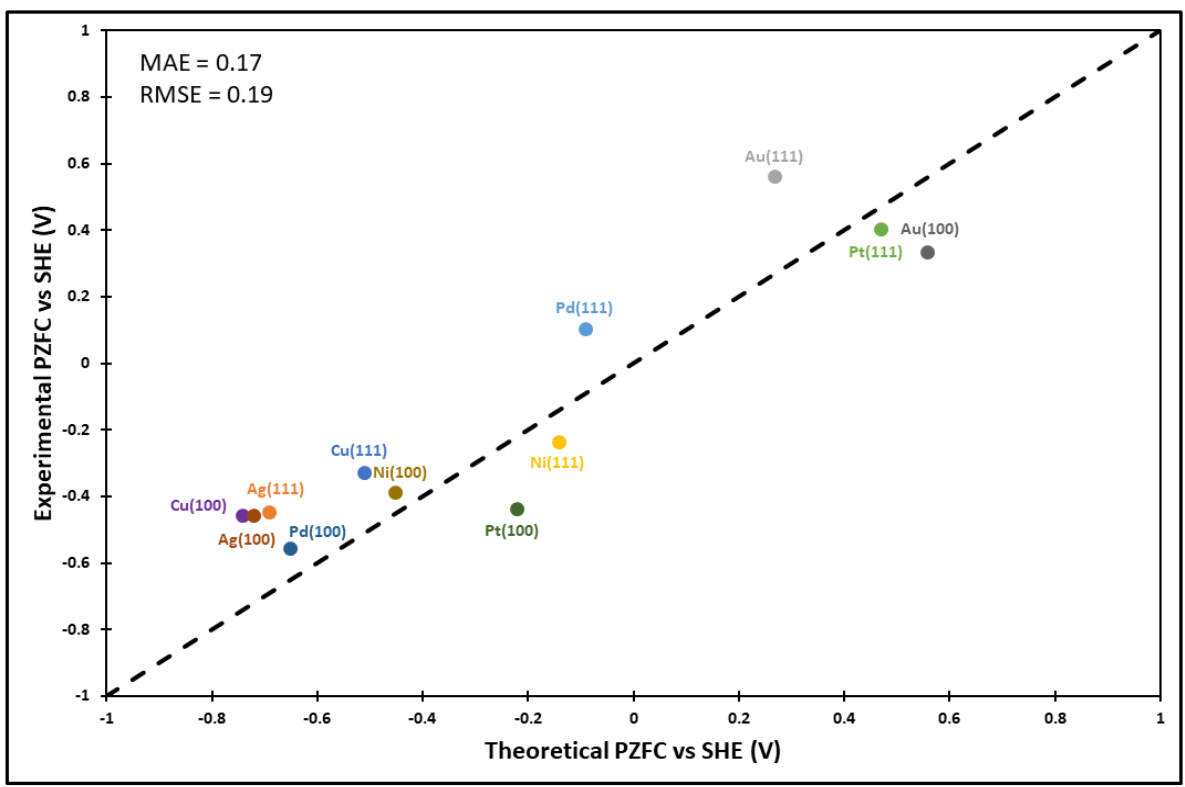

**Figure 1**. Comparing theory and experimental PZFC values.

From **Fig. 1c** and **Fig. 5**, it can be found that the PZFC values of the metal (111) surface are higher than that of the corresponding (100) surface. It suggests that the (100) surface behaves differently from the flatter (111) surface [49]. This matches the trend of the work function values of these metal surfaces [30]. This is because the PZFC values are derived from the work function of the clean surface covered by the ion-free water form [20]. The findings from previous studies have shown that the (100) surface of a material is more active due to the presence of low-coordination atoms [50, 51] (see **Fig. 6**). As a result, the electrons can be more difficult to be removed from the (111) surface, as evidenced by their higher work function in comparison with that of the (100) surface [30]. Indeed, the Au surfaces in this study showed an exception compared to the other metal surfaces. The DFT results using the hybrid solvation model revealed that the Au(111) surface exhibited lower PZFC values. This discrepancy can potentially be attributed to the reconstruction of the Au(111) surfaces [52], which was not taken into account in our unreconstructed Au(111) surface model used in the calculations.

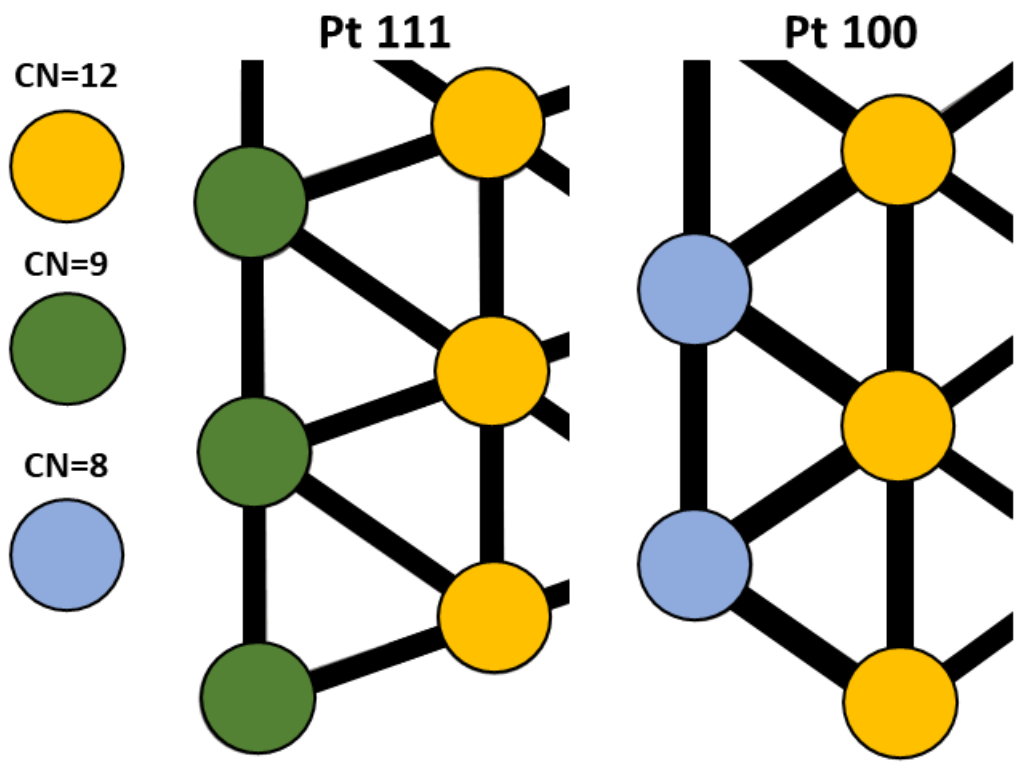

**Figure 6**. Coordination numbers of atoms in the topmost layer of Group 10 and 11 *fcc* metal (111) and (110) surfaces. CN stands for coordination number.

The discrepancy between the PZFC values using the implicit and hybrid solvation models suggests that water adsorption can have a considerable impact on the PZFC values while the interactions between water molecules and the metal surface are relatively weak due to the large distances between atoms. Our simulations in the study revealed that the M-O distance between water molecules and the metal surface never decreased below 3 Å. To determine the role of the adsorption strength of water on PZFC values, we calculated the $E_{ad}$ was calculated as:

$$E_{ad} = \frac{E_{total} - E_{Clean} - E_{H2O}*n}{n} \tag{3}$$

where $E_{total}$ is the total energy of the explicit only system, $E_{clean}$ is the energy without the implicit or explicit methods, $E_{H2O}$ is the energy of a single water molecule and n is the number of waters in the model. All the adsorption energies of water molecules onto the metal surface are higher than -1 eV. In addition, the trend in $E_{ad}$ of Au≈Ag<Cu<Pd<Pt<Ni is observed in other works [53]. This trend is consistent on the (111) and (100) surfaces. Especially, the results showed that Au, Ag, and Cu, known as hydrophobic metals [37], exhibited weaker interactions with water compared to the other metals studied.

The relationship between the PZFC and the adsorption properties, e.g., $d_{M-O}$ and $E_{ad}$, are shown in **Fig. 7**. It can be found that the adsorption properties of water have a weak correlation to the PZFC. It indicates that the orientations of water molecules are a more important factor influencing PZFC values. Interestingly, the water orientations vary for the different metals. Pt(111)-water interface exists in DOWN configuration, which is also supported by some recent studies [19, 28, 43, 54]. MIX configuration was found to be more popular. Of all 12 metal surfaces, the MIX water configuration is the most stable one on seven surfaces. This is because the MIX configuration combines characteristics from both UP and DOWN counterparts, making it a versatile choice on most of the metal surfaces. Especially, MIX configuration was energetically preferred when metal-water interactions are weak, as the water molecules can easily rotate. Meanwhile, UP and DOWN water configurations were preferred on one and four metal surfaces, respectively. The only surface with UP configuration is Cu(100), which has the smallest PZFC value among all the considered surfaces. The success of the hybrid solvation method in accurately determining the PZFC can be attributed to the optimized water orientation, which plays a vital role in achieving realistic predictions of electrochemical properties at the solid-liquid interface. Neglecting the presence of water can lead to significant deviations between predicted and actual behaviors of the electrified interface. In this study, the inclusion of water molecules in the model provided a more accurate representation of the system, leading to improved agreement with experimental observations.

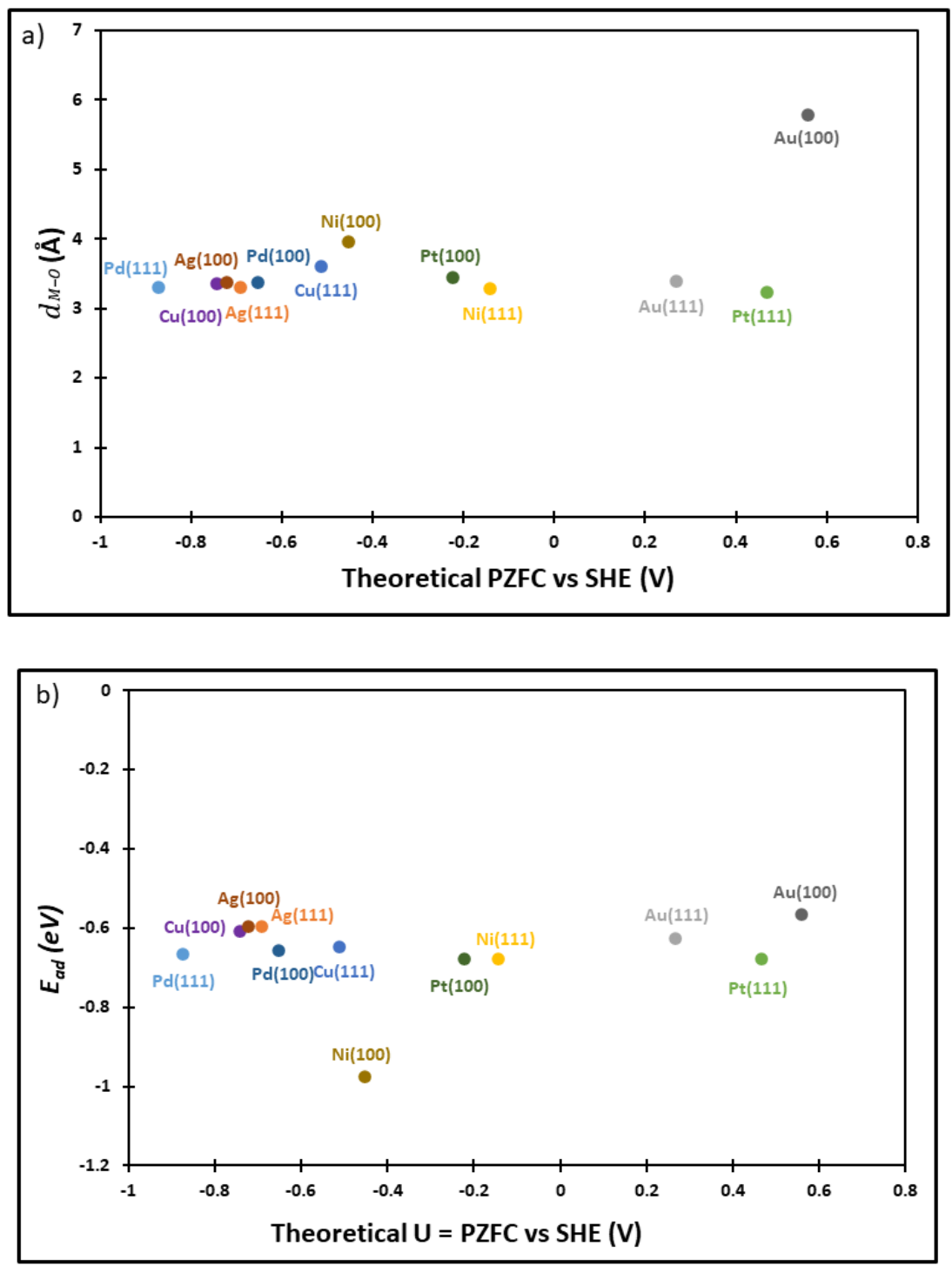

**Figure 7.** a) The relationship between the interfacial $d_{M-O}$ distance and the theoretical PZFC values; b) The relationship between the water adsorption energy and the theoretical PZFC values.

## 4. Conclusion

In this study, we employed DFT method to determine the PZFC values for Group 10 and 11 metal (111) and (100) surfaces. The PZFC value is a critical parameter to understand the solid-liquid interface, but its experimental and theoretical determination has historically been challenging. Our computational results revealed that DFT calculations using an implicit solvation model were inadequate in generating accurate PZFC values for these metal surfaces. To overcome this limitation, we introduced a hybrid solvation model, combining one layer of explicit water molecules adjacent to the surface with an implicit electrostatic continuum. This hybrid approach successfully reproduced the experimental PZFC values for the metal surfaces, demonstrating the importance of including explicit water layers in PZFC computations. Furthermore, our study identified that both the surface structures and the orientation of water molecules play significant roles in determining the PZFC values. This finding highlights the crucial influence of water on the electrochemical behavior at the solid-liquid interface. By providing a cost-effective and accurate theoretical method, our research opens new avenues for further theoretical investigation on the PZFC values of solid-liquid interfaces. This advancement has substantial implications for understanding material properties and optimizing electrochemical processes, paving the way for more efficient and sustainable energy technologies.

**CrediT Authorship Contribution Statement**

**Jack Jon Hinsch**: Investigation, Visualization, Conceptualization, Methodology, Formal Analysis, Data curation, Writing – original draft, Validating. **Jessica Jein White**: Methodology. **Yun Wang**: Supervision, Resources, Funding acquisition, Writing – review & editing

**Conflict of Competing Interest**

The authors declare that they have no known competing financial interests or personal relationships that could have appeared to influence the work reported in this paper.

**Declaration Data Availability**

All data is available in manuscript and supplementary text.

**Declaration Acknowledgement**

This research was conducted on the supercomputers in National Computational Infrastructure (NCI) in Canberra, Australia, which is supported by the Australian Commonwealth Government, and Pawsey Supercomputing Centre in Perth with the funding from the Australian government and the Government of Western Australia. We acknowledge funding from CNRS Institute of Chemistry through the "International Emerging Actions 2022" mobility grant (2DH2 project). The authors acknowledge financial support from the Australian Research Council Discovery Project (Grant No. DP210103266).